\documentclass[final,5p,times,twocolumn,authoryear]{elsarticle}

\usepackage{amssymb}
\usepackage{amsthm}
\usepackage{amsmath}
\journal{Planetary and Space Science}

\begin{document}

\begin{frontmatter}

\title{Imaging polarimetry and spectropolarimetry of comet C/2013 R1 (LoveJoy)\tnoteref{titlenote}}
\tnotetext[titlenote]{Based on data collected with 2m RCC telescope at Rozhen National Astronomical Observatory.}

%% \author{Galin B. Borisov\fnref{myfootnote}}
%% \address{Radarweg 29, Amsterdam}
%% \fntext[myfootnote]{Since 1880.}

\author[ia_address,armagh_address]{Galin Borisov\corref{correspondingauthor}}
\cortext[correspondingauthor]{Corresponding author}
\ead{gborisov@astro.bas.bg}

\author[armagh_address]{Stefano Bagnulo}

\author[ia_address]{Plamen Nikolov}

\author[ia_address]{Tanyu Bonev}

\address[ia_address]{Institute of Astronomy and National Astronomical Observatory, 
Bulgarian Academy of Sciences, 72, Tsarigradsko Chaussee Blvd., BG-1784 Sofia, Bulgaria}
\address[armagh_address]{Armagh Observatory, College Hill, Armagh BT61 9DG, Northern Ireland, UK}

\begin{abstract}

%% Comet C/2013 R1 (Lovejoy) was observed during a multi-instrument campaign with 
%% the 2m Ritchey-Chr\'etien-Coud\'e (RCC) telescope of the Bulgarian National 
%% Astronomical Observatory (BNAO) Rozhen from 20 Dec 2013 until 07 Jan 2014. 
%% Imaging and spectropolarimetric data were obtained on December 29 and January 3, 
%% respectively, with the 2-Channel-Focal-Reducer Rozhen (FoReRo2) \citep{Jockers2000} 
%% attached at the Cassegrain focus of the telescope. In the polarimetric mode, FoReRo2 
%% is equipped with a Wollaston prism, placed before a dichroic beam splitter, that splits 
%% the signals into two different channels, allowing us to re-construct polarimetric maps 
%% of extended objects in two spectral regions simultaneously, using narrow-band filters. 
%% By replacing the filters with two grisms we can perform spectropolarimetric
%% measurements. All our measurements were obtained using the beam-swapping technique.

We have obtained imaging polarimetry of the comet C/2013 R1 (Lovejoy) with 2-Channel-Focal-Reducer Rozhen 
instrument at 2m Ritchey-Chr\'etien-Coud\'e telescope of the Bulgarian National 
Astronomical Observatory Rozhen in two dust continuum filters covering wavelength 
intervals clear from molecular emissions and centred at 4430\,\AA\,\, in (blue filter) and at 6840\,\AA\,\, in (red filter). 
In imaging mode we measured the degree of linear polarisation $17.01\pm0.09$\,\% in the blue and $18.81\pm0.02$\,\% in the red, 
which is in a very good agreement with measurements of other comets at the similar phase angle. 
We have also obtained polarisation maps in both filters. We found a strong correlation between the spatial 
distribution of the polarisation and the dust colour. Spectropolarimetry of the 
nucleus region shows an increase of the polarisation with wavelength, 
and a depolarisation in the spectral regions with gas emission lines, most noticeable 
in C$_2$ emission band, which shows a polarisation of $6.0\pm1.1$\,\%. 

%We will compare depolarisation 
%of the molecules with the expected theoretical value of 1/7 from Feofilov \cite{Feofilov1961}.
%Spectropolarimetry of the nucleus region was found consistent with narrow-band filter 
%polarimetry, i.e., the polarisation increases with wavelength. 

\end{abstract}

\begin{keyword}
comets \sep C/2013 R1 (Lovejoy) \sep polarimetry \sep spectropolarimetry \sep dust \sep molecules
\end{keyword}

\end{frontmatter}

%%\linenumbers

\section{Introduction}

Polarimetry is sensitive to the physical properties 
of the dust particles: size, shape, porosity, orientation and chemical composition 
represented by its material complex refractive index. Polarimetric measurements 
give us the possibility to determine some parameters that cannot be 
determined trough traditional intensity measurements.

The first polarimetric observations of comets were made by Fran\c{c}ois 
\citet{Arago}, who discovered the polarised light in the Great Comet 1819 II.

Later, observations of comets clarified some common characteristics of 
the polarised light, for example that usually the plane of polarisation is 
perpendicular to the scattering plane, and that there are variation of the 
polarisation in different parts of the comet (coma, tail). Contemporary 
polarimetric observations of comets began with the work of Yngve \citet{Ohman39,Ohman41}, 
who observed for the first time the continuum polarisation 
in comets and the polarisation of the emission lines.

Most of the recent polarimetric observations of comets have been obtained by  
Kiselev and collaborators. \citet{CometsPolDB} has also created a database with 
more than 2600 measurements of linear and circular polarisation for 
64 comets since 1940s. 

Most of the polarimetric observations of small Solar system bodies are aimed at measuring 
the variation of the polarisation with phase angle (which actually is 
$180^\circ - $ scattering angle) and also its dependence on wavelength.  

From the theoretical side, many works have been carried out by Kolokolova and 
collaborators \citep{Kol97Ic,Kol97PSS,KolCometsII}.

The polarisation of the dust jet-like structures in the dust coma of the comet Hale--Bopp 
was obtained for the first time by \citet{Hadamcik97} and was discussed later on in \citet{Hadamcik2003}.

A recent review of all comets investigation can be found in the books by \citet{SSbook} and \cite{Kiselev2015}. 

Comet C/2013 R1 (Lovejoy) was discovered by Terry Lovejoy (Thornlands, Queensland, 
Australia) with images acquired on 2013 September 7 and 8, using his 20-cm 
reflector and a CCD camera.

Other polarimetric measurements of the comet C/2013 R1 (Lovejoy) are presented by 
\citet{Furusho2014} (imagine polarisation with the Subaru telescope) and by 
\citet{Rosenbush2014} (linear and circular polarimetric measurements and their modelling).

\section{Observations}

Comet C/2013 R1 (Lovejoy) was observed during a multi-instrument campaign 
with the 2m Ritchey-Chr\'etien-Coud\'e (RCC) telescope of the Bulgarian 
National Astronomical Observatory (BNAO) Rozhen from 20 Dec 2013 until 07 Jan 2014. 
Because of the target brightness we could achieve a relatively high S/N ratio 
and obtain high accuracy of polarimetric measurements. C/2013 R1 (Lovejoy) was a 
new comet which approach to the inner Solar System for the first time and giving us an 
opportunity to investigate the pristine material from the era of the Solar System formation.

\subsection{Instrumentation} 

\begin{figure}[!ht]
\begin{center}
\includegraphics[width=\columnwidth]{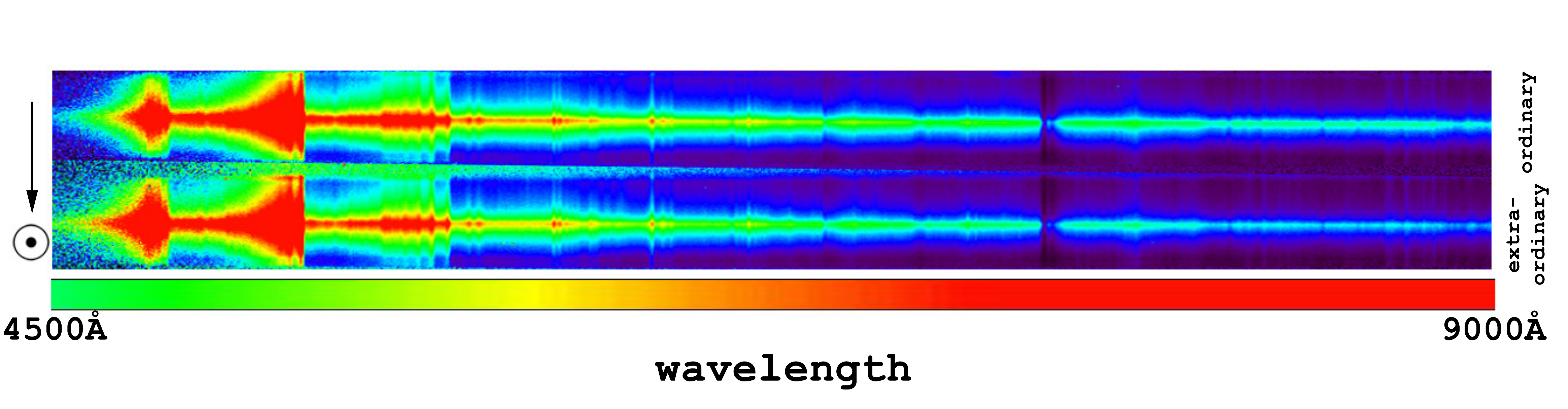}
\caption{Spectropolarimetric image of comet C/2013 R1 (Lovejoy).}
\label{sppolimg}
\end{center}
\end{figure}

Polarimetric observations were performed with the 2-Channel-Focal-Reducer Rozhen (FoReRo2) 
\citep{Jockers2000} attached at the Cassegrain focus of the 2m RCC telescope. In polarimetric 
mode, FoReRo2 is equipped with a Wollaston prism, placed before a dichroic beam splitter, which 
splits the signals into two different channels, allowing us to re-construct polarimetric maps of 
extended objects in two spectral regions simultaneously, using narrow band filters. By replacing the 
filters with two grisms, we can perform spectropolarimetric measurements. 
An example of raw spectropolarimetric image can be seen in Fig.~\ref{sppolimg}.

\begin{figure}[!ht]
\begin{center}
\includegraphics[width=0.65\columnwidth, angle=90]{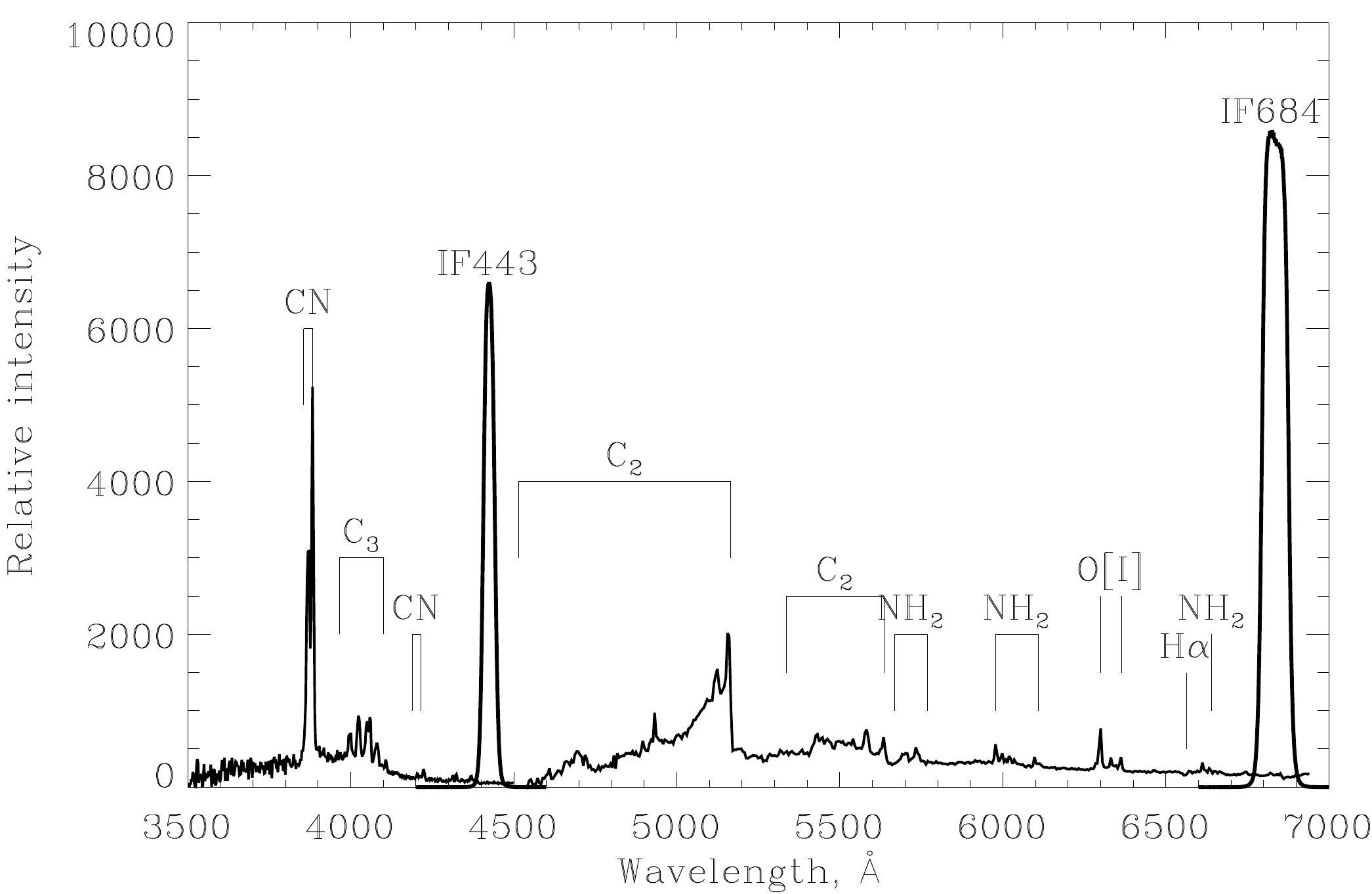}
\caption{Continuum filters transmission curves (IF443 \& IF684) overplotted on a comet spectrum}
\label{filters}
\end{center}
\end{figure}

Imaging polarimetry was 
obtained in two dust continuum filters covering wavelength intervals clear from molecular emission 
and centred at 4430\,\AA\,\,and 6840\,\AA, having a passband of 35\,\AA, and 71\,\AA\,\,and hereafter called 
IF443 and IF684, respectively (see Fig.~\ref{filters}).

\subsection{Comet C/2013 R1 (Lovejoy)}

Imaging and spectropolarimetric data were obtained on December 29 and January 3 
respectively, with FoReRo2. The geometrical conditions during the observations 
are shown in table~\ref{obs}.

\begin{table}[!ht]
\begin{minipage}{\columnwidth}
\begin{center}
\begin{tabular}{c|c|c|c|l}
\hline
Date              & r\footnote{Heliocentric distance},\,AU        & $\Delta$\footnote{Geocentric distance},\,AU & $\alpha$\footnote{Phase angle},\,deg & Obs. mode                            \\
\hline
\hline
      20 Dec 2013  &       0.8132    & 0.8562         &      72.2    &        HRS\footnote{High resolution spectroscopy}         \\
      21 Dec 2013  &       0.8123    & 0.8765         &      71.1    &        HRS         \\ 
      22 Dec 2013  &       0.8118    & 0.8965         &      70.1    &        HRS         \\
      23 Dec 2013  &       0.8118    & 0.9161         &      69.1    &        HRS         \\
      24 Dec 2013  &       0.8122    & 0.9355         &      68.1    &        HRS         \\
{\bf 29 Dec 2013} & {\bf 0.8210}  & {\bf 1.0305}  & {\bf 63.0}  & {\bf ImPol}\footnote{Imaging polarimetry}                 \\
      30 Dec 2013  &       0.8240    & 1.0495         &      62.0    &        NBF\footnote{Gas and dust coma imaging in Narrow band filters}            \\
      31 Dec 2013  &       0.8275    & 1.0675         &      61.0    &        H$_2$O$^+$              \\
{\bf 03 Jan 2014} & {\bf 0.8405}  & {\bf 1.1205}  & {\bf 58.1}  & {\bf SPol}\footnote{Spectropolarimetry}                  \\
      08 Jan 2014  &       0.8631    & 1.1865         &      54.6    &        NBF \& H$_2$O$^+$  \\
\hline
\end{tabular}
\caption{Observing Log}
\label{obs}
\end{center}
\end{minipage}
\end{table}

\subsection{Data Reduction}
\label{subsec:BS}

All images were pre-processed through a standard bias subtraction and
flat field correction.

At the time of our observations, the polarimetric optics of FoReRo2
included a Wollaston prism but not a retarder waveplate, preventing us
from adopting a beam-swapping technique to minimise instrumental
effects \citep[see, e.g.][]{Bagnulo2009}. Previous experience
showed that polarimetric observations with FoReRo2 were affected by
non-negligible and non-constant instrumental polarisation.  In an
attempt to mitigate this problem, we decided to obtain observations at
two instrument position angles, one with the principal plan of the
Wollaston prism aligned to the scattering plan (i.e., the plan defined
by the sun, the comet and the observer) and one perpendicular to
it. By denoting with $f^\parallel$ and $f^\perp$ the fluxes in the
parallel and in the perpendicular beams, and with $k_\parallel$ and
$k_\perp$ the transmission functions in the parallel and in the
perpendicular beam of the Wollaston prism respectively, the observed quantity
is
%%%%%%%%%%%%%%%%%%%%%%%%%%%%%%%%%%%%%%%%%%%%%%%%%%%%%%%%%%%%%%%%%%%%%%%%%%%%%%%%%%%%%%%%%%%
\begin{eqnarray}
%\begin{array}{rcl}
\hat{\frac{Q}{I}} &=& 
\frac{1}{2}\left[\left(\frac{f^\parallel-f^\perp}{f^\parallel+f^\perp}\right)_{{\rm PA}=\phi+90^\circ}-
                 \left(\frac{f^\parallel-f^\perp}{f^\parallel+f^\perp}\right)_{{\rm PA}=\phi} \right] \nonumber \\ 
                 \nonumber \\ 
                  &=&
\frac{1}{2}\left[\frac{k_\parallel(I+Q)-k_\perp(I-Q)}{k_\parallel(I+Q)+k_\perp(I-Q)}-
                 \frac{k_\parallel(I-Q)-k_\perp(I+Q)}{k_\parallel(I-Q)+k_\perp(I+Q)} \right] \nonumber\\
                 \nonumber \\ 
                  &=& 
                 \frac{(k_\parallel+k_\perp)^2IQ-(k_\parallel-k_\perp)^2\,IQ}{(k_\parallel+k_\perp)^2I^2-(k_\parallel-k_\perp)^2\,Q^2} \nonumber
%\end{array}
\end{eqnarray}
%%%%%%%%%%%%%%%%%%%%%%%%%%%%%%%%%%%%%%%%%%%%%%%%%%%%%%%%%%%%%%%%%%%%%%%%%%%%%%%%%%%%%%%%%%%
where $\phi$ is the angle is the angle between the direction Object-North Pole and the direction Object-Sun.
If $k_\parallel \simeq k_\perp$ we obtain
\begin{equation}
\frac{Q}{I} \simeq \hat{\frac{Q}{I}}\;. \nonumber
\end{equation} 
where $Q/I$ is the reduced Stokes parameter $Q$ measured assuming as a reference direction that one 
perpendicular to the scattering plane. We did not measure $U/I$, assuming that for symmetric reasons 
it is probably zero. Of course, the images were combined together after a 90 degrees rotation.

\section{Results}

\subsection{Aperture polarimetry}

\begin{figure}[!ht]
\begin{center}
\includegraphics[width=\columnwidth]{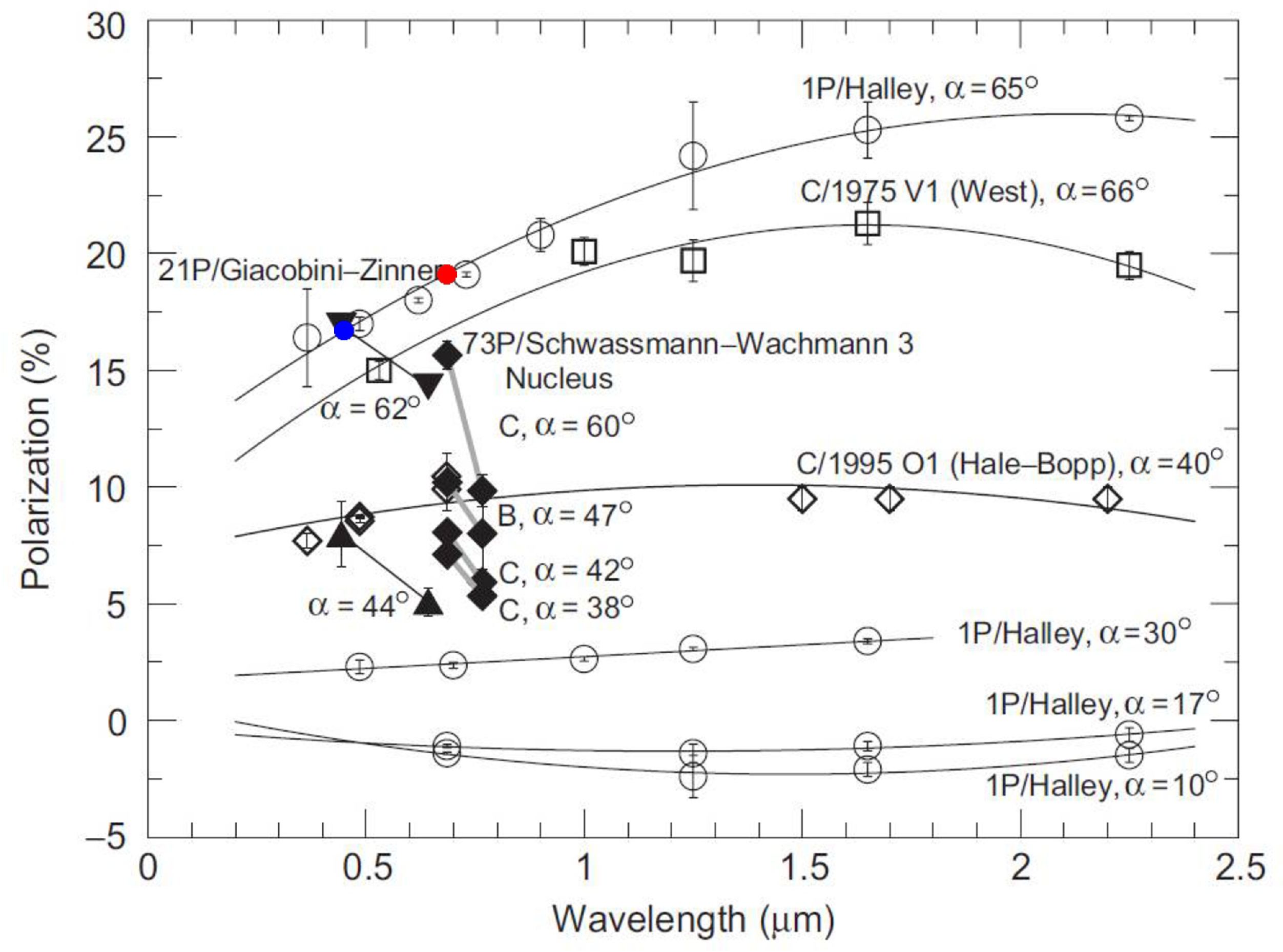}
\caption{Comparison of polarimetric measurement of comet C/2013 R1 (Lovejoy) 
and Kiselev database \citep{CometsPolDB} (error bars of our measurements are smaller than the symbols)}
\label{LJ_pol}
\end{center}
\end{figure}

The aperture photometry with a circular aperture with a radius of $13\times10^3$\,km on the comet 
was performed to measure the intensity of the two orthogonal polarised beams 
$f^\parallel$ and $f^\perp$. The background was measured in the farthest possible 
place in the sunwards direction. The results for both continuum filter 
IF443 and IF684 are $P_{443}=17.01 \pm 0.09 \%$ and  $P_{684}=18.81 \pm 0.02 \%$ 
respectively. Their comparison with the Kiselev database \citep{CometsPolDB} is 
shown in Fig.~\ref{LJ_pol} and is in a good agreement with the data for the comet 1P/Halley 
at similar phase angles (65$^{\circ}$).

\subsection{Imaging polarimetry}

In order to investigate the spatial distribution of the linear polarisation of the 
light reflected from cometary dust we performed imaging polarimetry of the comet's coma.

\begin{figure}[!ht]
\begin{center}
\includegraphics[width=\columnwidth]{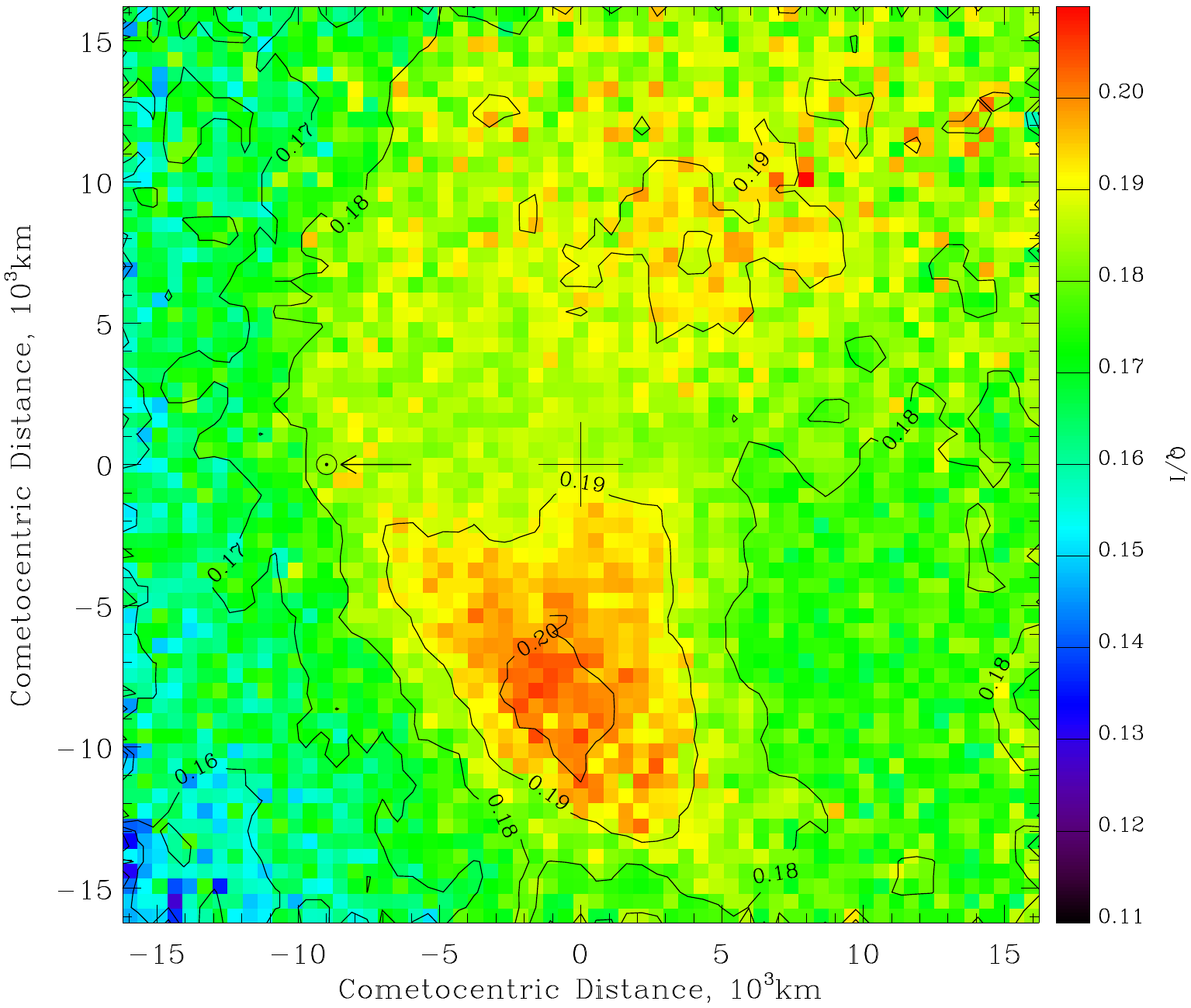}
\includegraphics[width=0.88\columnwidth]{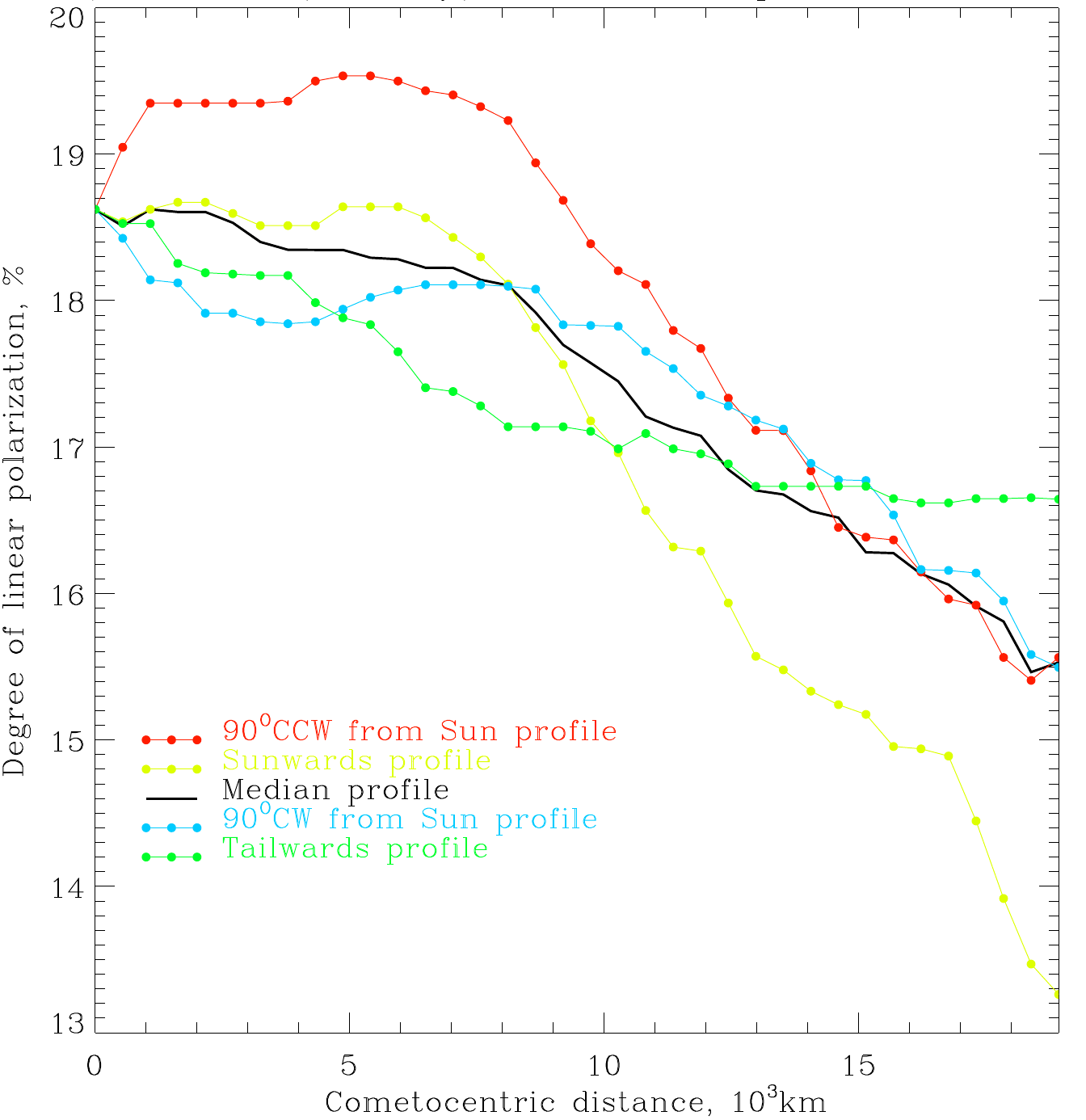}
\caption{Polarisation map with a marked comet photocentre ($+$) 
and direction to the Sun ($\odot\leftarrow$) (top panel) 
and radial profiles in different directions (bottom panel) in IF684}
\label{polmaps}
\end{center}
\end{figure}

We construct polarisation maps by calculating the degree of linear 
polarisation, using the beam swapping technique, for each pixel 
in a pre-selected region around comet photo centre. The resultant polarimetric 
maps and radial profiles for the dust continuum filter IF684 are presented 
in the top and bottom panel of Fig.~\ref{polmaps}, respectively.

Due to the low signal-to-noise ratio, the polarimetric map in IF443 dose not display any remarkable structures. 
By contrast, the polarisation map in IF684 shows clearly a highly polarised structure apart from the nucleus. 

The polarisation degree in a square region 11$\times$11\,$10^3$\,km enveloped this structure 
is slightly higher than that of the total coma, i.e. $P^{\rm jet}_{684}=18.83 \pm 0.045\,\%$

\begin{figure}[!ht]
\begin{center}
\includegraphics[width=\columnwidth]{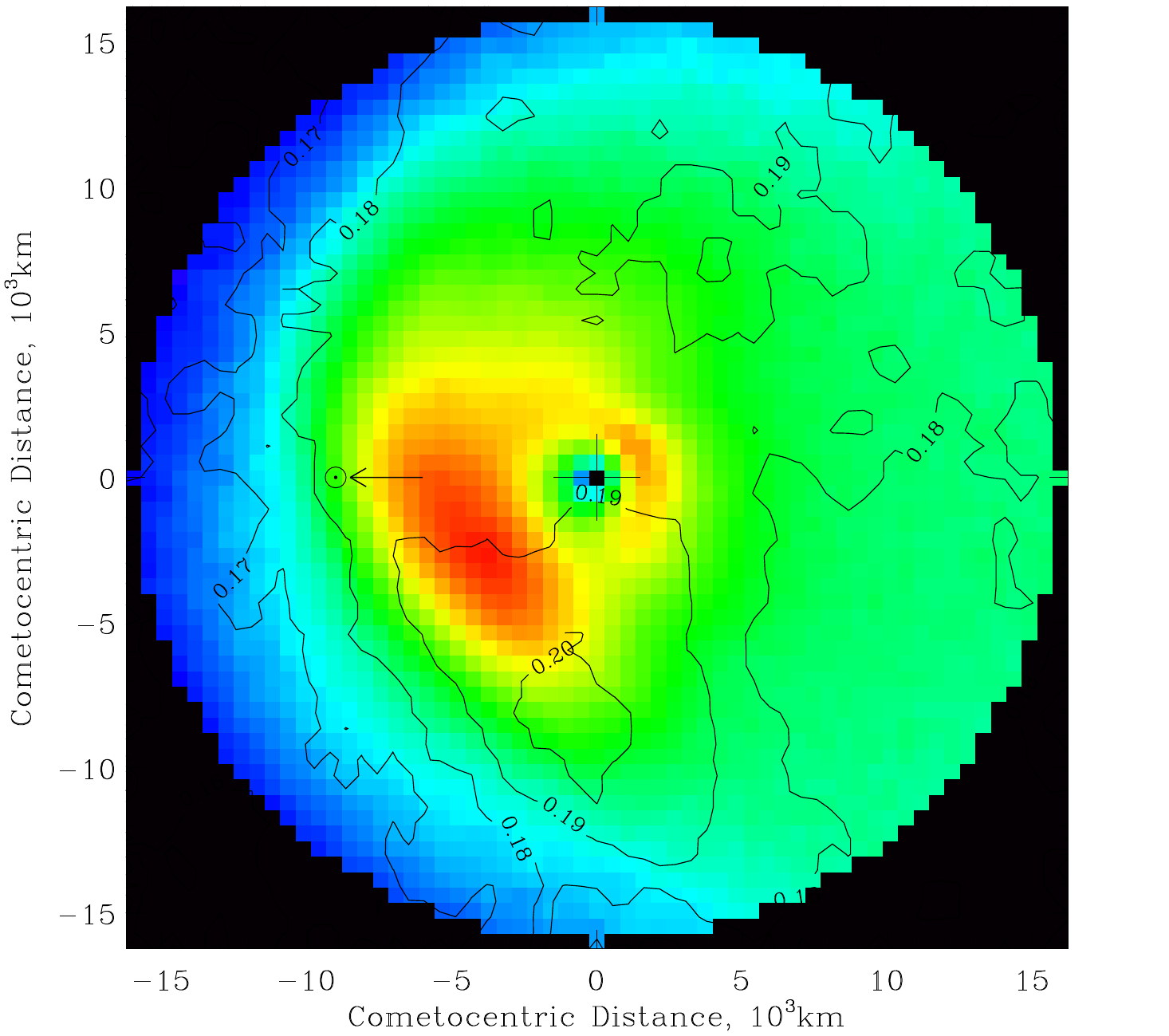}
\includegraphics[width=\columnwidth]{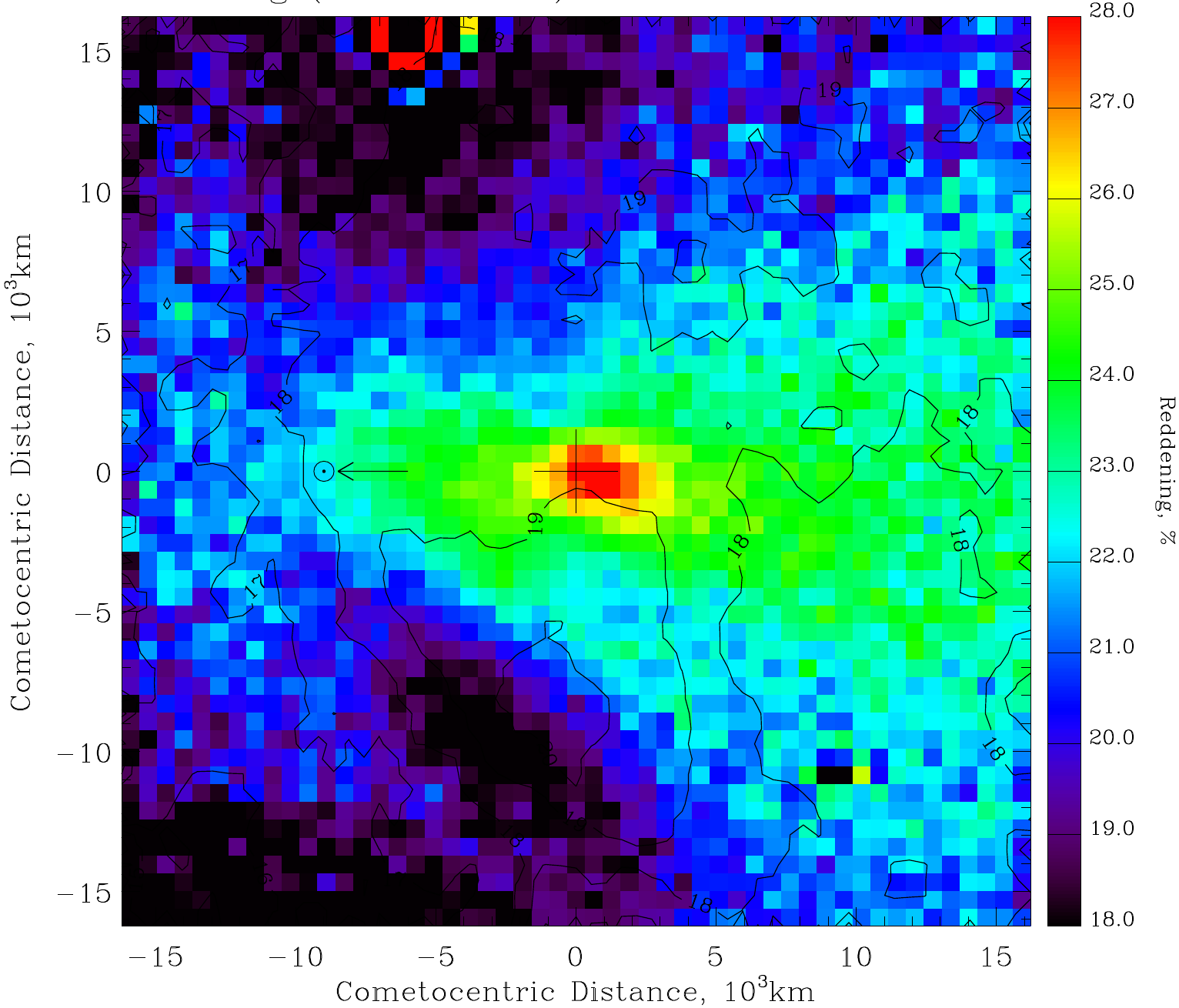}
\caption{Enhanced coma structure in IF684 (top panel) and colour map (bottom panel) 
compared with polarisation (overplayed contours)}
\label{col-pol}
\end{center}
\end{figure}

In order to investigate the nature of this 
structure, we first try to connect this highly polarised structure with 
a jet-like structure in the dust coma. Therefore we used an enhanced 
procedure to reveal any inhomogeneity around the nucleus caused by 
unsteady outflow from the nucleus. This procedure remove the average coma, 
modelled by fitting the radial profiles of the dust at different azimuths with a power low. 
Afterwards this model was subtracted and only the jet-like structure remains in the image. 
In the top panel of Fig.~\ref{col-pol}, such a structure is clearly visible, 
but it is shifted from the polarisation one, presented with overplotted contours.

Next, we compared it with the distribution of the dust normalised 
reflectivity gradient, or so called reddening. That is why we construct the 
so called colour map of the dust by calculating the quantity 

\begin{gather}
S'(\lambda_1,\lambda_2)=\frac{1}{\bar{S}}\frac{\partial S}{\partial\lambda}, \nonumber
\label{eq:color}
\end{gather}
where $\bar{S}=(S_{\lambda_{2}}-S_{\lambda_{1}})/2$ and is usually expressed in $\%/1000\AA$, for each 
pixel in both dust continuum images in IF443 and IF684. The resultant colour 
map is presented in the bottom panel of figure~\ref{col-pol}. The connection 
between high polarisation and low reddening is clearly visible. The interpretation 
is that low reddening means small particles, or very dark ones, both of which 
give high polarisation.

\subsection{Spectropolarimetry}

\begin{figure}[!ht]
\begin{center}
\includegraphics[angle=90, width=\columnwidth]{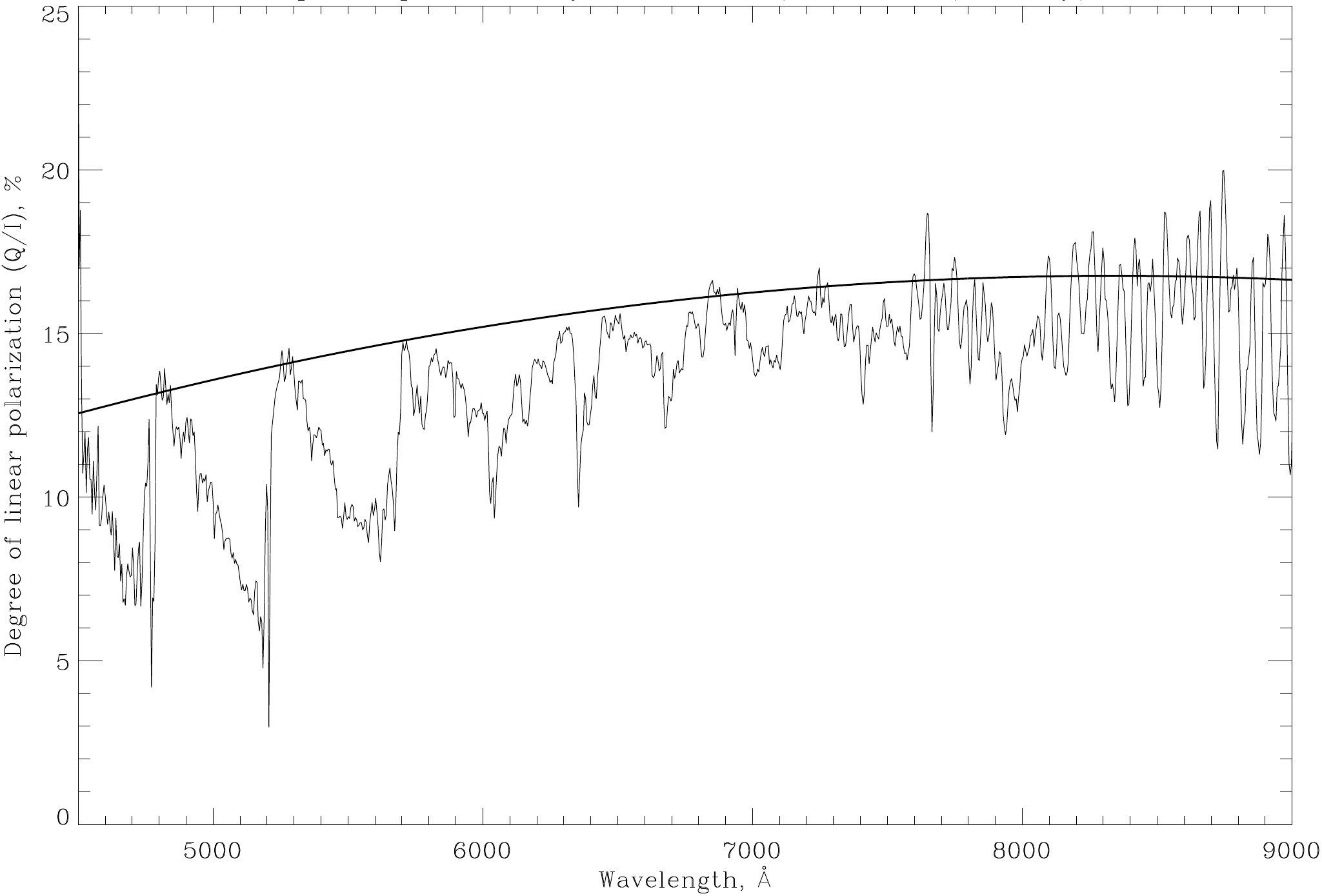}
\caption{Spectropolarisation of comet C/2013 R1 (Lovejoy) and 
a fitted continuum polarisation.}
\label{sppol}
\end{center}
\end{figure}

Polarisation spectrum of the comet C/2013 R1 (Lovejoy) is shown in Fig.~\ref{sppol}.

\begin{figure}[!ht]
\begin{center}
\includegraphics[angle=90, width=\columnwidth]{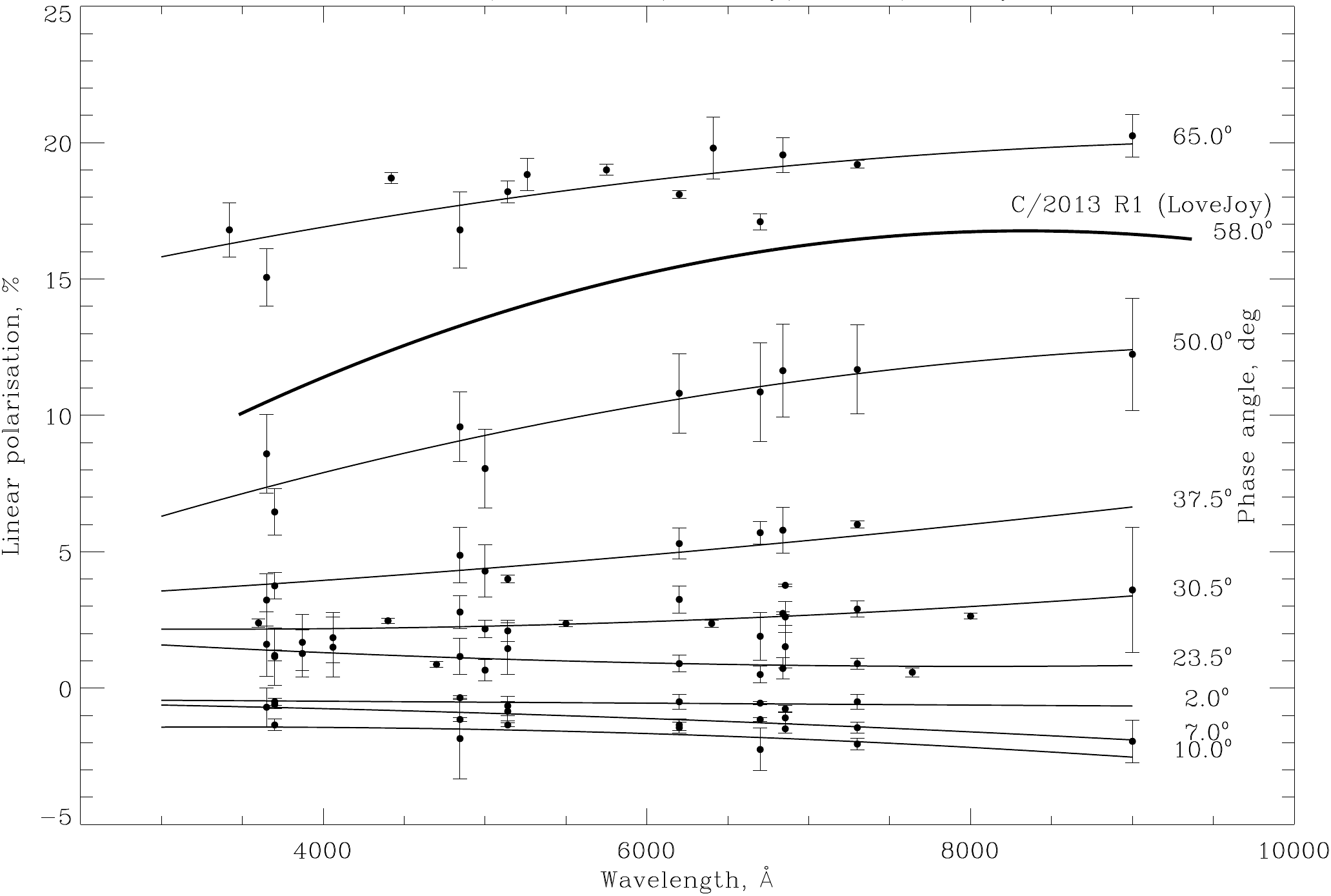}
\caption{Comparison of the comet C/2013 R1 (LoveJoy) with the typical comet 1P/Halley 
measurements at different phase angles.}
\label{LJ1P}
\end{center}
\end{figure}

We carried out the the polynomial fit to the dust continuum polarisation. 
The result for comet 1P/Halley at different phase angles are also presented 
\citep{CometsPolDB} and our observations are very well fitted to the phase angle 
and wavelength trend of the comet 1/P Halley (see Fig.~\ref{LJ1P}).

\begin{figure}[!ht]
\begin{center}
\includegraphics[width=\columnwidth]{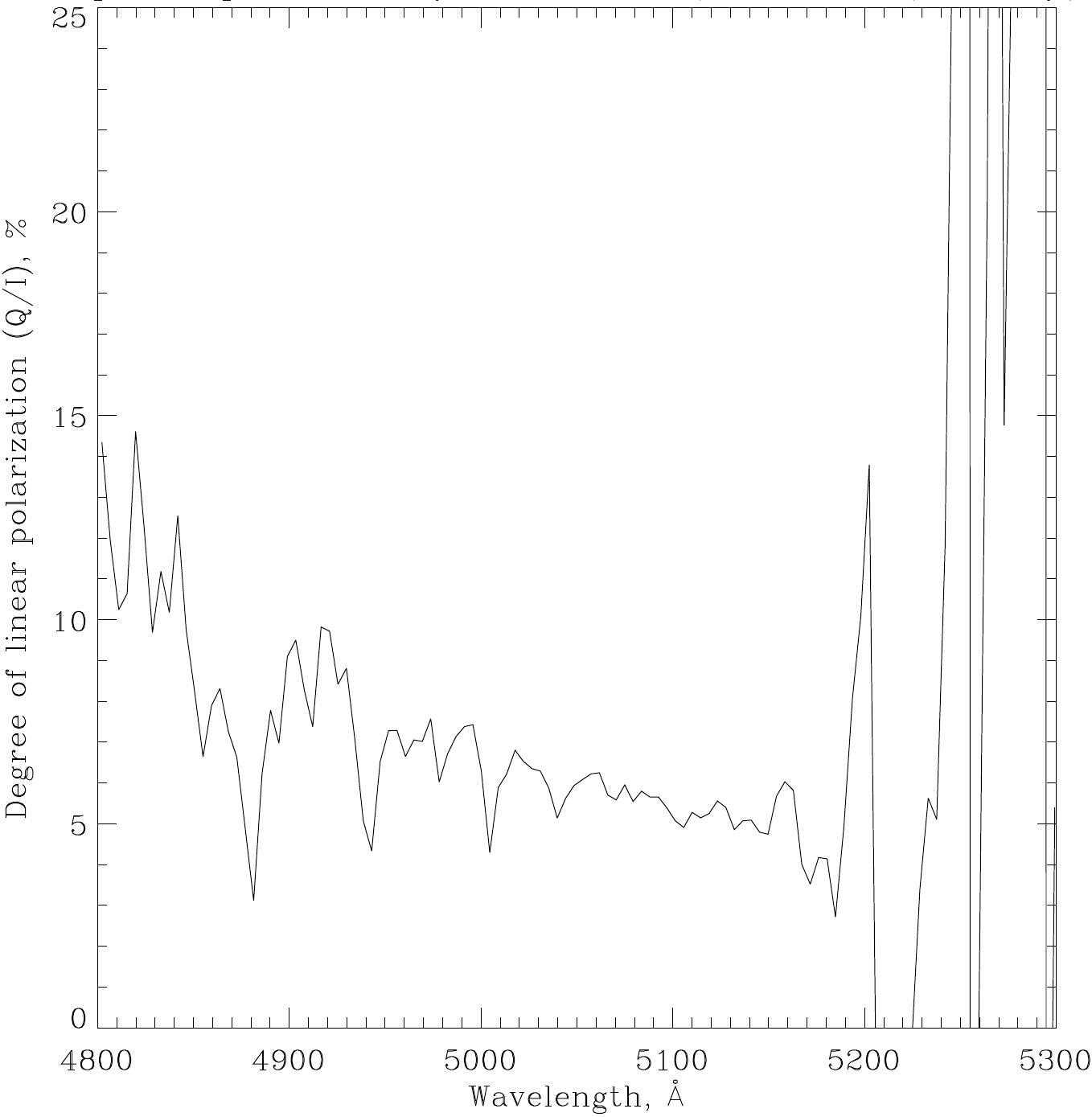}
\caption{Polarisation spectrum of the C$_2$ emission line in the coma of the comet C/2013 R1 (LoveJoy)}
\label{c2pol}
\end{center}
\end{figure}

The polarisation in molecular spectral lines is lower than that in the continuum. 
According to the theory of diatomic molecules it should be a constant of $1/7$ \citep{Feofilov1961}. 
To estimate the polarisation of the emission lines we first subtract the continuum contribution from 
both spectra in the ordinary and extraordinary beams. Then, using the remaining 
flux from resonance fluorescence of the molecules, we calculate the 
degree of linear polarisation following the procedure described in section~\ref{subsec:BS} 
for each wavelength bin. The result for the main C$_2$ molecular band is 
presented in Fig.~\ref{c2pol}.

The average value of linear polarisation over the whole molecular band of C$_2$ 
is $6.0\pm1.1$\,\%. This value is less than half that of the theoretical one, 
which is based on theory of anisotropic rotating oscillators. 
Also these values are for pure gas without the presence of other species. The theory 
says that polarisation of the fluorescence of diatomic molecules is reduced when 
foreign gases are introduced, and collisions occur, resulting a  
depolarisation. Therefore we conclude that this lower value of linear polarisation 
in the C$_2$ molecular band is due to collisions with other species in the comet's 
coma. 

The excitation by unidirectional natural radiation leading to linear polarisation
of an emission line has a maximum polarisation value for a phase angle of 90$^{\circ}$. 
The expected variation of gas polarisation with phase angle $\alpha$ is given by the 
expression:

\begin{gather}
P(\alpha)=\frac{P_{\rm max}\sin^2\alpha}{1+P_{\rm max}\cos^2\alpha} \nonumber
\end{gather}

From our measurements we calculate P$_{\rm max}=8.5\substack{+2.5 \\ -1.6}$\,\%. 
\citet{Swamy} says that the theoretical value of P$_{\rm max}$ for the C$_2$ 
and CN molecules is 7.7\,\%, which is within our confidence interval {6.9 -- 11.2}. 

\section{Conclusions}
\begin{itemize}
\item We have measured the degree of linear polarisation in the continuum filters IF443 and IF684 
			of the dust ejected from comet C/2013 R1 (Lovejoy), and our results  
			({$P_{443}=17.01 \pm 0.09 \%$} and {$P_{684}=18.81 \pm 0.02 \%$}) 
			are in a good agreement with measurements of the typical comet 1P/Halley..
\item We have investigated the spatial distribution of the dust in both continuum filters. 
			A polarisation feature in the red was found and compared with jet-like 
			structures in the dust coma, and connected with low values in the map of the 
			dust colour, which shows that this region is populated rather with small or very dark grains.
\item \citet{Furusho2014} presented imaging polarisation of the comet Lovejoy with 
			the Subaru telescope, and their results (including the detection of coma features with high polarisation values) 
			are consistent with our results.
\item We have obtained a spectropolarimetry of the comet and found the change of the degree 
			of linear polarisation of the dust with wavelength. This is in a good agreement with 
			the results obtained for comet 1P/Halley. 
\item In the C$_2$ molecule emission band we have measured $6.0\pm1.1$\,\% degree of linear polarisation. 
			The results deviate from the theoretical value for diatomic molecules 14.3\,\% 
			\citep{Feofilov1961}, but are in good agreement with a value measured for other comets 
			P$_{\rm max}$=7.7\,\% \citep{Swamy}. This can be used to explain the depolarisation effect in 
			the molecular coma.
\end{itemize}

\section*{Acknowledgements}

The authors gratefully acknowledge the observing grant support from the Institute of 
Astronomy and Rozhen National Astronomical Observatory, Bulgarian Academy of 
Sciences.

GB and SB also gratefully acknowledge financial support from the COST Action MP1104 
"Polarisation as a tool to study the Solar System and beyond".

GB also gratefully acknowledge financial support from the Federation 
of Finnish Learned Societies and the organisers of the ACM2014.

PN acknowledge financial support from ESF and Bulgarian Ministry of Education and 
Science under the contract BG051PO001-3.3.06-0047.

\bibliography{GBorisov-ACM2014-PSS}

\end{document}